\documentclass[11pt]{article}
\usepackage{graphicx}
\usepackage{bm}
\usepackage{hyphenat}

\title{Reply to Comment on ``Acoustically assisted spin-transfer-torque switching 
of nanomagnets: An energy-efficient hybrid writing scheme for non-volatile memory'' [Appl. Phys. Lett., 103, 232401 (2013)]}
\author{Ayan K. Biswas$^1$, Jayasimha Atulasimha$^2$ and Supriyo Bandyopadhyay$^1$ \\
\\
$^1$Department of Electrical and Computer Engineering, \\
Virginia Commonwealth University, 
Richmond, Virginia 23284, USA \\
$^2$Department of Mechanical and Nuclear Engineering, \\
Virginia Commonwealth University,
Richmond, Virginia 23284, USA}
\date{}

\begin{document}

\maketitle

A comment \cite{kuntal} has been posted in arXiv on our work on energy-efficient acoustically-assisted spin-transfer-torque writing of bits in
non-volatile magnetic memory \cite{ayan1}. It raises the following points: (1)
our acoustically-assisted spin-transfer-torque random access memory 
(AA-STT-RAM) \cite{ayan1} is less energy-efficient than transistors, (2) our Terfenol-D based AA-STT-RAM is 
less energy-efficient than 
a CoFeB based traditional STT-RAM, (3) the surface acoustic wave in our AA-STT-RAM
causes some small magnetization rotation even in cells that are not being written into and that 
results in 40 kT of standby (static) energy dissipation per cell. This wasted energy makes our AA-STT-RAM memory cell 
worse than a transistor in energy dissipation, (4)
we under-estimated the energy barrier 
in our elliptical nanomagnet (memory cell) by $\sim$33\% because we used a formula that is valid only when the 
eccentricity of the ellipse is small, whereas our eccentricity was 0.57 (major axis $a$ = 110 nm; minor 
axis $b$ = 90 nm, eccentricity = $\sqrt{1 - (b/a)^2}$), (5) an AA-STT-RAM cell's footprint is larger than that of a transistor and hence our
memory will be less dense than transistor-based memory, and (6) the commenter's ``own idea'' of 
writing a bit in a non-volatile magnetic memory cell with the aid of strain alone 
and a sensing element \cite{kuntal2} is somehow more ``attractive'' than our acoustically assisted 
spin transfer torque based writing scheme. 

We reply as follows: Point (1) is specious. One does not write bits into a single 
transistor since it is {\it volatile}. One should therefore compare our AA-STT-RAM 
(which is non-volatile) with a transistor-based non-volatile memory
cell, e.g. a NAND flash. That dissipates several orders of magnitude more energy than an AA-STT-RAM and has several 
orders of magnitude worse endurance as well \cite{paper}. There is no comparison between the two.

Point (2) is equally specious. Comparing a Terfenol-D based AA-STT-RAM with a CoFeB based STT-RAM is comparing
an apple to an orange. To compare apples with apples, one should compare a Terfenol-D based AA-STT-RAM with a 
Terfenol-D based STT-RAM, or a CoFeB based AA-STT-RAM with a CoFeB based STT-RAM, or choose any other material
as long as it is the same for both RAMs. We carried out the analysis for Terfenol-D and found the AA-STT-RAM to be 
superior to STT-RAM in energy dissipation.
Another group carried out the analysis for CoFe independently and they too found that strain-assisted STT-RAM
(which is equivalent to AA-STT-RAM) is superior \cite{Intel}. We have now carried out the analysis for CoFeB and once again found the 
AA-STT-RAM to be superior. So far, whenever apples have been compared with apples, the AA-STT-RAM has eclipsed
the STT-RAM.

Point (3) is incorrect. Let us concede for the sake of argument that the SAW wave dissipates 40 kT in each 
cell of an AA-STT-RAM per cycle of $\sim$ 1 ns duration. The resulting power dissipation is 0.16 nW. This is power dissipated 
in the cell when it is {\it not} being written into and hence should be compared to 
the leakage power (i.e. static power or standby power) dissipation in transistors. For the 22-nm node transistor technology, 
the leakage current, 
or off-current, is $\sim$ 0.7 nA in a 1 $\mu$m gate width transistor for a power supply voltage of 1.5 V (threshold voltage 
300 mV) 
and 0.8 $\mu$A for a power supply voltage of 0.5 V (threshold voltage 100 mV) \cite{iwai}. The corresponding leakage powers are 1.05 nW
and 400 nW, respectively. Therefore, the standby power dissipation in an AA-STT-RAM cell is almost an order
of magnitude {\it less} than that 
in a single transistor, let alone a NAND flash cell made of multiple transistors.

Point (4) is somewhat orthogonal to the message of our paper. If we rigorously calculate the energy
barrier from Ref. \cite{paper2} (which is valid for small or large eccentricities), then we will find that the energy 
barrier we used actually corresponds
to a major axis dimension of 108 nm (instead of 110 nm) and a minor axis dimension of 93 nm (instead of 90 nm).
Since lithographic precision of 2-3 nm is somewhat impractical when delineating the nanomagnets, the dimensions
are rounded off to 110 nm and 90 nm. In any case, if the energy barrier increases, then the energy dissipation 
in {\it both} STT-RAM and AA-STT-RAM will go up, whereas if the energy barrier decreases, the energy dissipation 
in {\it both} STT-RAM and AA-STT-RAM will go down. Since both types of RAMs are affected similarly, the comparison between the two 
(which is the message of the paper) and the conclusion that AA-STT-RAM is superior, is not altered

Point (5) is a red herring. It is well known that magnets are larger than transistors and cannot
be shrunk beyond the super-paramagnetic limit at the operating temperature. This limitation is more than
offset by the non-volatility, endurance, etc. of magnets that transistors do not possess.

Point (6) is orthogonal to our paper. It refers to an idea (that we actually co-authored but did not pursue 
further; it is certainly not
the commenter's sole idea) whereby
a bit is written into a non-volatile magnetic memory cell with strain alone, without any spin polarized current 
generating a spin transfer torque. On the surface, it may appear to be a more energy-efficient strategy
for writing bits than switching with both
strain and STT (since strain consumes much less energy than STT to rotate a magnet's magnetization), but 
its debilitating drawback is that 
it requires a feedback/sensing circuit to write bits correctly (i.e. with better than 50\% error probability).
When the energy dissipated in the feedback circuit is factored in, that scheme may not be any more 
energy-efficient than acoustically-assisted spin-transfer-torque \cite{Comment} and almost surely 
will be more error-prone since the feedback circuit needs to operate with very precise timing. Fortunately, there
are better ways to write bits in magnetic memory with strain alone \cite{pertsev,Tiercelin,ayan2,ayan,penn}. They may indeed dissipate 
less energy than acoustically-assisted spin-transfer-torque, but one disadvantage some of them have is that the 
writing step must be preceded by a reading step every time. That disadvantage also afflicts the 
scheme in ref. \cite{kuntal2}. However, 
the acoustically-assisted spin-transfer-torque technique is free of this disadvantage.

In conclusion, we find that none of the points raised in Ref. \cite{kuntal} is tenable.

\end{document}